\documentclass{aastex62}

\graphicspath{{./}{figures/}}

\received{July , 2019}
\revised{xxx, 2019}
\accepted{xxx,xxx} %\today
\submitjournal{ApJL}

\shorttitle{Synchrotron emission from ordered magnetic fields.}
\shortauthors{Sharma V. et al.}
\begin{document}

\title{Time varying polarized gamma-rays from GRB 160821A: evidence for ordered magnetic fields}

\correspondingauthor{ Vidushi Sharma, Shabnam Iyyani}
\email{vidushi@iucaa.in, shabnam@iucaa.in}

\author[0000-0002-4394-4138]{Vidushi Sharma}
\affil{Inter-University Center for Astronomy and Astrophysics, Pune, Maharashtra 411007, India}

\author[0000-0002-2525-3464]{Shabnam Iyyani}
\affiliation{Inter-University Center for Astronomy and Astrophysics, Pune, Maharashtra 411007, India}

\author{Dipankar Bhattacharya}
\affiliation{Inter-University Center for Astronomy and Astrophysics, Pune, Maharashtra 411007, India}

\author{Tanmoy Chattopadhyay}
\affiliation{Department of Physics, Stanford University, 382 Via Pueblo Mall,
Stanford CA 94305}
\affiliation{Kavli Institute of Astrophysics and Cosmology, 452 Lomita Mall,
Stanford, CA 94305}

\author{A. R. Rao}
\affiliation{Inter-University Center for Astronomy and Astrophysics, Pune, Maharashtra 411007, India}
\affiliation{Tata Institute of Fundamental Research, Mumbai, Maharashtra 400005, India}

\author{ E. Aarthy}
\affiliation{Physical Research Laboratory, Ahmedabad, Gujarat 380009, India}

\author{ Santosh V. Vadawale}
\affiliation{Physical Research Laboratory, Ahmedabad, Gujarat 380009, India}

\author{ N. P. S. Mithun}
\affiliation{Physical Research Laboratory, Ahmedabad, Gujarat 380009, India}

\author{ Varun. B. Bhalerao}
\affiliation{Indian Institute of Technology Bombay, Mumbai, India}

\author{ Felix Ryde}
\affiliation{Department of Physics, KTH Royal Institute of Technology, AlbaNova, SE-106 91 Stockholm, Sweden}
\affiliation{The Oskar Klein Centre for Cosmoparticle Physics, AlbaNova, SE-106 91 Stockholm, Sweden}

\author{Asaf Pe'er}
\affiliation{Department of Physics, University College Cork, Cork, Ireland}
\affiliation{Department of Physics, Bar-Ilan University, Ramat-Gan, 52900, Israel}



\begin{abstract}
GRB 160821A is the third most energetic gamma ray burst observed by the {\it Fermi} gamma-ray space telescope. Based on the observations made by Cadmium Zinc Telluride Imager (CZTI) on board {\it AstroSat}, here we report the most conclusive evidence to date of (i) high linear polarization ($66^{+26}_{-27} \%$; $5.3\, \sigma$ detection), and (ii) variation of its polarization angle with time happening twice during the rise and decay phase of the burst at $3.5\, \sigma$ and $3.1\, \sigma$ detections respectively. All confidence levels are reported for two parameters of interest. These observations strongly suggest synchrotron radiation produced in magnetic field lines which are highly ordered on angular scales of $1/\Gamma$, where $\Gamma $ is the Lorentz factor of the outflow.

\end{abstract}



\section{Introduction} \label{sec:intro} 
Gamma-ray bursts (GRBs) are the most intense astrophysical outbursts in the Universe. In the last several decades, the spectra of prompt gamma-ray emission of GRBs have been extensively studied using various space observatories like BATSE onboard CGRO \citep{Fishman2013}, {\it Niel Gehrels Swift} \citep{Gehrels2004}, {\it Fermi} \citep{Meegan2009,Atwood2009} etc. The radiation process producing the prompt gamma-ray emission, however, still remains a mystery. Polarization along with spectrum measurements can provide an insight into this long standing enigma. Polarization measurement of prompt emission is highly challenging, mainly because of the scarcity of incident photons and the brevity of the event. Previously, polarization measurements of prompt gamma ray emission were attempted only for a few cases by RHESSI  \citep{Coburn_boggs_2003,Wigger_etal_2004}, INTEGRAL \citep{McGlynn2009,Gotz_etal_2009,Gotz_etal_2013,Gotz_etal_2014}, GAP \citep{Yonetoku_etal_2011, Yonetoku_etal_2012}, CZTI \citep{Rao_etal_2016,Chattopadhyay_etal_2017,Chand_etal_2018,Chand_etal_2019}, POLAR \citep{Zhang_etal_2019,Burgess_etal_2019} etc but the results were statistically less significant and sometimes unconvincing (for a recent review please refer  \citealt{McConnell2017}). In this Letter, for the first time, we present a conclusive evidence of polarization across the GRB 160821A in the energy range $100-300$ keV using CZTI (Cadmium Zinc Telluride Imager) instrument aboard {\it AstroSat} \citep{Singh_etal_2014}.  

On August 21, 2016 the Burst alert telescope (BAT)  on board {\it Neil Gehrels Swift} observatory \citep{Gehrels2004}, triggered and located the GRB 160821A  at (RA, Dec) = (171.248, 42.343) with 3 arcmin uncertainty \citep{BATrefined_GCN_2016} at 20:34:30 UT, along with other space observatories such as 
Konus-wind and the Gamma-Ray Burst monitor on board CALET. However, due to solar observing constraints {\it Swift} could not slew to the BAT position until a week. Hence, there was no X-ray Telescope (XRT) and UV/Optical Telescope (UVOT) observations for this burst. Half an hour after the trigger time, an optical transient was detected by ground based telescopes but no redshift measurement could be made. 

\begin{figure}
\plotone{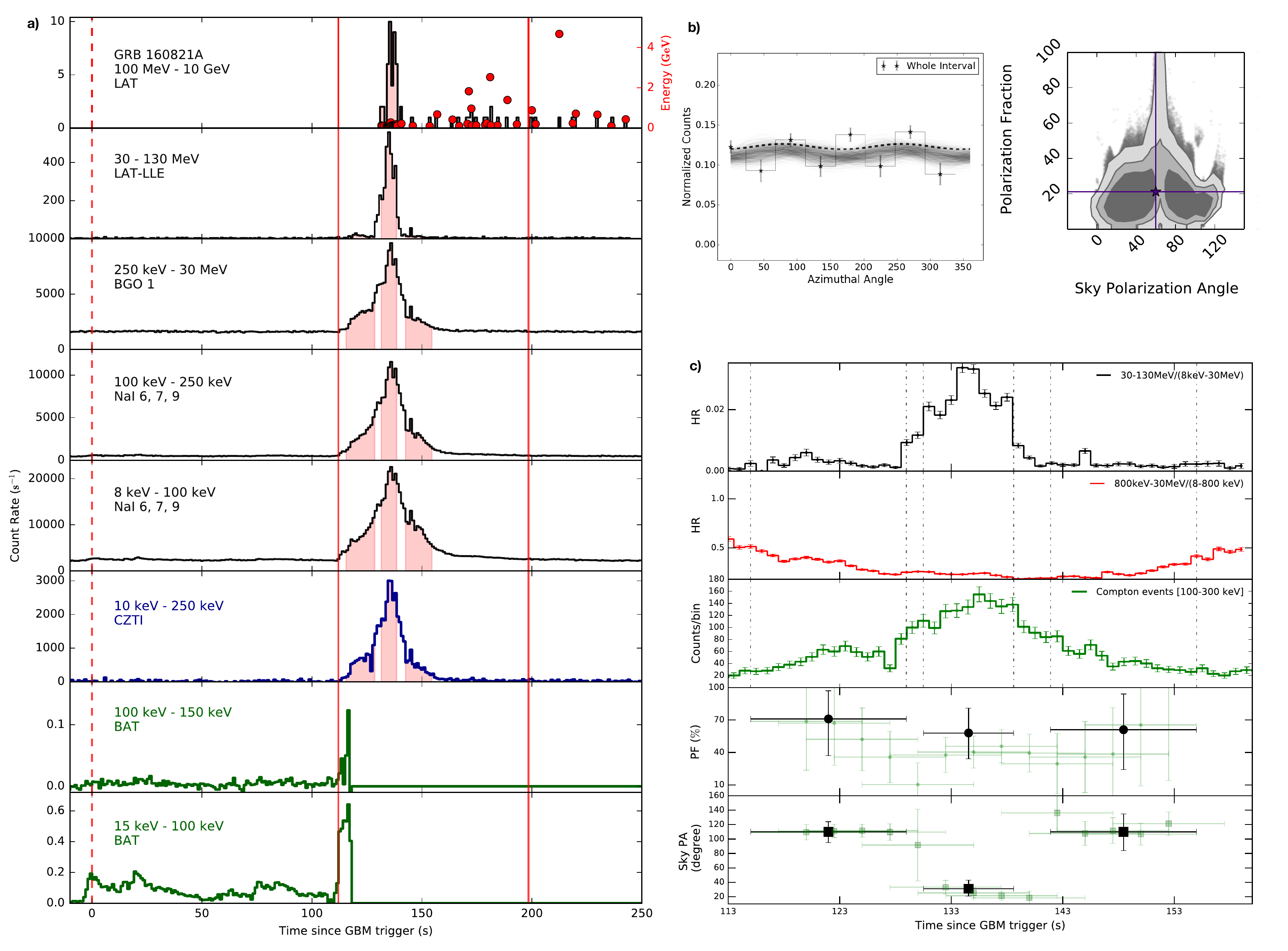}
\caption{a) A composite 1 s binned light curve of the burst is shown for {\it Fermi} detectors: LAT, LAT-LLE, BGO and NaI, {\it AstroSat}/ CZTI and {\it Swift}/ BAT. The red vertical dashed line (at $T_{0} = 0$ s) corresponds to the trigger time of {\it Fermi} and the red solid vertical lines mark the beginning and end of the main episode of the burst. The {\it Fermi}/LAT photons whose probability of association with the same GRB is greater than 90$\%$, are shown as red points in the upper panel with energy information scaling on the right side of y-axis. The time intervals studied in polarization analysis are shown in shaded red regions.
b) On the left, the azimuthal distribution and the best fit modulation (dashed line) obtained for the entire main episode of the burst are shown. Modulation fits obtained for 1000 runs out of the $1.1 \times 10^7$ simulation datasets are shown in the shaded black colour. On the right, the 2D histogram plot of the obtained PF and PA values along with the contours corresponding to confidence levels of  $68 \, \%, 90 \, \%, 99 \,\%$ are shown in darker to lighter colors of black which are over plotted on the scatter plot of PF and PA. The average value of PF and PA are marked by the violet star.
c) Uppermost and second top panels show the hardness ratio (HR) of the counts LLE to sum of the counts in NaI 6 and BGO 1 (black), and ratio of the counts in BGO 1 to that of NaI 6 respectively. 
In the third panel, the 1s binned CZTI Compton events (green) light curve is shown.
The time intervals (black vertical dash dot lines) for which the temporal polarization study is done, are shown. Fourth and fifth panels show the polarization fraction (PF, black circles) and corresponding polarization angle (PA, black squares) obtained for these time intervals. Also, PF and PA values obtained in the fine time resolved analysis are shown in the background in shaded green circles and squares respectively. 
 }
\label{lc_whole_pol}
\end{figure}

 {\it Fermi} Gamma-Ray Burst Monitor (GBM, \citealt{Meegan2009})  also triggered on the burst at 20:34:30.04 UT \citep{FermiGBM_GCN_2016}. The GBM light curve included a precursor emission starting from trigger time, $T_{0}$ till $T_{0} \,+\, 112\, \rm s$, followed by a bright emission episode with a duration T90\footnote{T90 for {\it Fermi} (or {\it AstroSat}) is defined as the time duration between the epochs when $5\%$ and $95\%$ of the total photon counts of the burst are detected in the energy range 50 -300 keV (40-200 keV).} 
 of $ 43$ s. 
 For the time interval $T_{0}\, -\, 4.1\,\rm s$ to $T_{0} \, + \,194.6\, \rm s$, an energy flux of $2.86 \pm 0.007 \times 10^{-6} \, \rm erg/cm^{2}/s$  
is obtained in 10-1000 keV band. This makes the burst the third brightest GRB observed by {\it Fermi} till date in terms of energy flux.
The {\it Fermi}-Large Area telescope (LAT, \citealt{Atwood2009}) detected emission in the LAT Low Energy (LLE) data ($30 - 100 $ MeV), starting at $ T_{0} + 116 \, \rm s $ and 
emission above $ 100$ MeV starting at $ T_{0} + 130 \, \rm s $ \citep{LATrefined_GCN_2016}, with LAT emission extending up to $\sim 2000\, \rm s$ beyond the duration of GBM emission. 
{\it AstroSat}-CZTI also detected the burst for a duration T90 of $42$ s \citep{Astrosat2016} and captured only the main episode of the burst. GRB 160821A was incident on CZTI from the direction, $\theta = 156.2^{\circ}$ and $\phi = 59.2^{\circ}$, thus coming through the side veto detector. 
Polarization measurement was attempted in the energy range 100 - 300 keV using $\sim 2549$ detected Compton events \citep{Chattopadhyay_etal_2014, Chattopadhyay_etal_2017}.  
In Figure \ref{lc_whole_pol}a, a composite 1 s binned light curve obtained from various detectors on board {\it Fermi}, {\it AstroSat} and {\it Swift} satellites are shown. 
The present work focuses only on the results of the study of the main episode of the burst observed by both {\it Fermi} and {\it AstroSat}. 

\section{Polarization analysis}
The measurement of polarization is obtained from the azimuthal distribution of Compton scattered photons, which lie preferentially in a direction orthogonal to the incident electric field vector \citep{Covino_Gotz2016,McConnell2017}. The azimuthal distribution is fitted with the cosine function of the form, 
\begin{equation}
C(\eta) = A \; cos(2(\eta - \phi_0 + \pi/2)) + B
\end{equation} 
where $\phi_0$ is the polarization angle (PA) of the incident photons as measured in the CZTI detector plane, $A/B$ is the modulation factor ($\mu$) and $\eta$ is the azimuthal angle (also see \citealt{Zhang_etal_2019}).
%, and their respective PF and PA values were determined. 
The measured detector plane PA is converted into the celestial/ sky reference frame by taking into account the satellite orientation at the time of observation and these values are reported throughout the Letter unless otherwise mentioned. The polarization fraction (PF) is calculated by normalizing the modulation with $\mu_{100}$ (the modulation factor obtained for $100\%$ polarized emission coming from the same direction of the GRB with the same spectrum) for the respective detector PA.

A low polarization (PF $= 21_{-19}^{+24}\, \%$) was found at $90\%$ ($2.2 \, \sigma$) confidence level for two parameters of interest, when the entire pulse constituting a time interval of 115 - 155 s was analyzed (Figure \ref{lc_whole_pol}b). All the errors quoted for polarization measurements in the Letter are at $68\%$ confidence interval of two parameters of interest including the systematic errors (Appendix \ref{sys_err}) unless otherwise mentioned.
%The observed modulation was better modelled using a constant line than a sinusoidal function (Bayes Factor\footnote{The Bayes factor (BF) obtained for null hypothesis model of unpolarized emission (constant line) to an alternative model of polarized emission (sinusoidal function). BF$<2$ favours the null hypothesis model. } = 0.7).  
We find this result to be in agreement with the recent polarization observations of GRBs made by POLAR \citep{Zhang_etal_2019}, where they find the time integrated emission of GRBs to possess a low polarization fraction. They suspect that such low polarization could be due to change in polarization angle within the pulses/ across different pulses. 

The evolution of the light curves observed by {\it Fermi} in different energy bands were characterized by studying the ratio of photon counts observed in high and low energies, parameterized as the hardness ratio (HR), see panels 1 and 2 in Figure \ref{lc_whole_pol}c. We found that the emission above $30 \, \rm MeV$ changed distinctly with respect to the rest of the burst after $T_{0} + 129\, \rm s$ and $T_{0} + 140\, \rm s$. In addition to this, a fine time resolved polarization analysis of the main episode was conducted. Polarization fraction and polarization angle for successive 10 s intervals shifted by 2.5 s were measured, thereby studying the evolutionary trend in PF and PA (Figure \ref{lc_whole_pol}c). Such a methodology including overlapping time intervals was adopted because of the limited number of photons available in the smaller time intervals. Therefore, during the analysis we tried to constrain the PF such that at least the lower limit of $68\%$ confidence interval of one parameter of interest was greater than zero. This is because if the time interval is unpolarized i.e PF is consistent with zero, then its PA has no physical meaning. 
A change in PA was observed to occur twice as the burst transited from its rise to peak and later to its decay phases while PF was greater than zero.  

We note that during the times when the PA angle makes a change, a decrease in PF is expected. 
Thus, based on the clear change in PA where we could constrain the polarization at a higher significance (i.e the lower limit of $99\%$ confidence interval of two parameters of interest of PF lie greater than zero) and the observed change in HR at high energies, we performed a relatively coarser time-resolved polarization analysis of the GRB by dividing the main episode into three non-overlapping time intervals: 115-129 s, 131 -139 s and 142 -155 s, which correspond to the rise, peak and decay phase of the burst respectively. 

%Based on these observations, for the temporal study of the observed polarization, we divided the main episode into three non-overlapping time intervals: 115-129 s, 131 -139 s and 142 -155 s, which correspond to the rise, peak and decay phase of the burst respectively, such that during the analysis, we can constrain the PF . 
%This division was based on the hardness ratio
%\footnote{The evolution of the light curves observed by {\it Fermi} in different energy bands were characterized by studying the ratio of photon counts observed in high and low energies, parameterized as the hardness ratio.} (HR) study of the {\it Fermi} light curves (see panels 1 and 2 in Figure \ref{lc_whole_pol}b) and the fine time resolved polarization analysis. 

\subsection{Results} 
\begin{figure}
\plotone{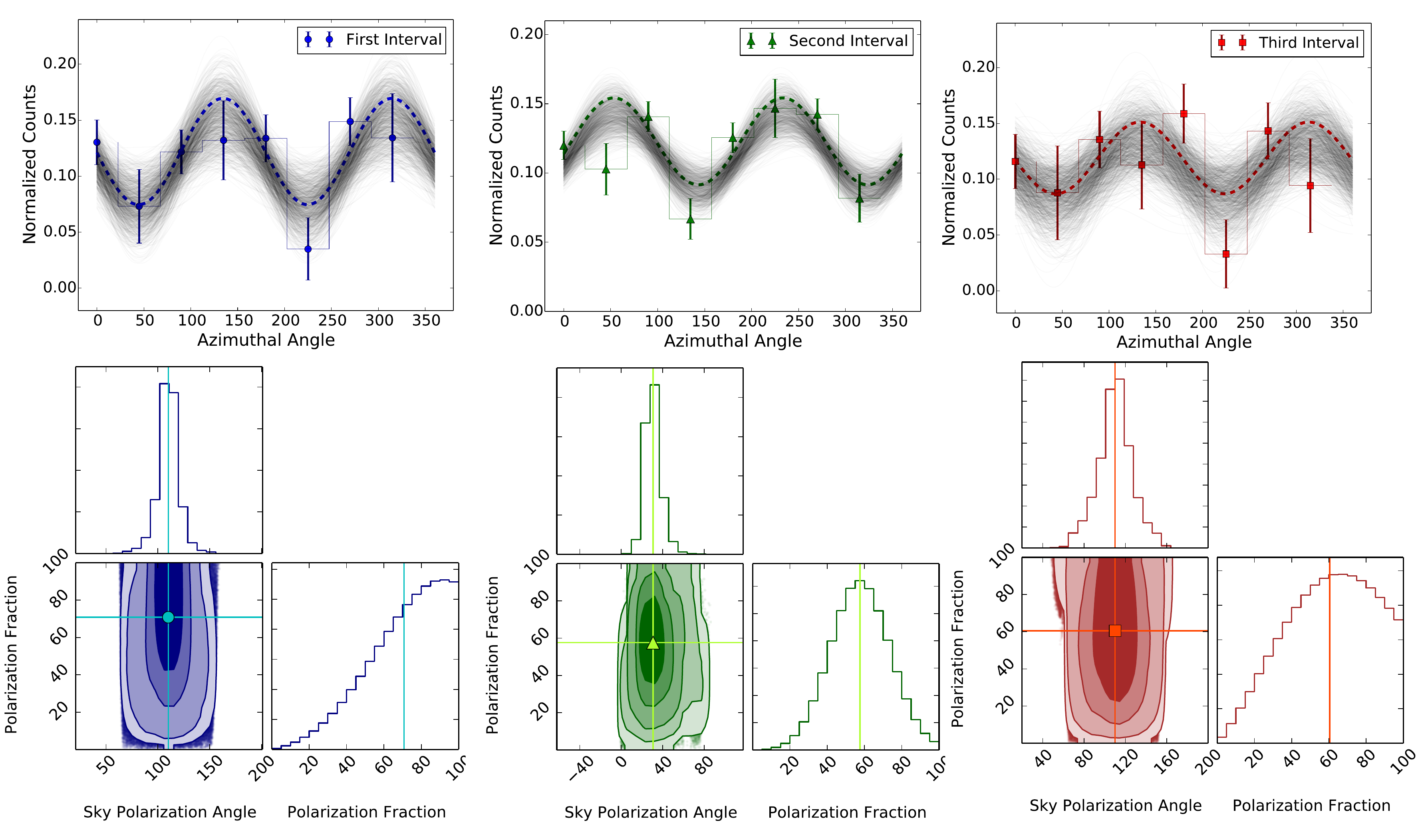}
\caption{In the left top corner, for the first interval, the best fit modulation curve obtained is shown in blue dashed line. Modulation fits obtained for 1000 out of $\sim 10^{7}$ simulated data sets are shown here in shaded black color. In the left bottom corner the 2D histogram plot of the obtained PF and PA values, along with the contours corresponding to confidence levels of $68 \, \%, 90 \, \%, 99 \,\%$ and $99.9\, \%$ are shown in terms of darker to lighter color shades of blue which are over-plotted on the scatter plot of PF and PA.  
Similar corresponding plots for the interval 2 are shown in the top middle and bottom middle (green colour). For this time interval, an additional contour of confidence level $99.99\,\%$ is shown in the 2D histogram plot.  For interval 3, corresponding plots are shown in the right top corner and right bottom corner (red colour). The average values of PF and PA obtained in the intervals 1,2,3 are marked by the blue circle, green triangle and red square respectively on their 2D histograms and they are marked by solid lines on their respective posterior distribution plots as well.}
\label{pol_results}
\end{figure}

%The time resolved polarization analysis finds that 
During the first time interval, the burst emission has a polarization fraction of $71^{+29}_{-41}\,\%$ and a polarization angle of $110^{+14\, \circ}_{-15}$. In the second interval, a PF = $58^{+29}_{-30}\%$ with a corresponding PA = $31^{+12\, \circ}_{-10}$ and in the third interval, a PF  = $61^{+39}_{-46}\, \%$ with a corresponding PA = $110^{+25\, \circ}_{-26}$, are determined (Figure \ref{lc_whole_pol}c). 
%All the quoted errors in the paper represent $68 \,\%$ confidence intervals for two parameters of interest including the systematic errors unless otherwise mentioned. 
By performing Monte Carlo simulations (Appendix \ref{posterior}) of the dataset of each interval, the posterior distributions of PF and PA of these intervals were also obtained. Intervals 1, 2 and 3 were found to be polarized at confidence levels of $99.8\, \%$ ($3.5\, \sigma$), $99.97\, \%$ ($4\, \sigma$) and $99.1\, \%$ ($3.1\, \sigma$) respectively for two parameters of interest (Figure \ref{pol_results}). 
As the burst transits from its rise to the peak phase and then into its decay phase, the PA shifts by $81^{\circ} \pm 13^{\circ}$ and $80^{\circ} \pm 19^{\circ}$ respectively (Figure \ref{lc_whole_pol}c). The statistical significance of the change in polarization angles, $\Delta \, PA_1$ and $\Delta \, PA_2$ are determined at $3.5 \, \sigma$ and $3.1\, \sigma$ respectively, which are the minimum of the obtained statistical significance of the two intervals to be polarized.  
%were obtained by doing Monte Carlo simulations involving combined fits of intervals 1, 2 and intervals 2, 3 respectively, whose PF paramters were linked while the polarization angles were kept free. In Figure \ref{avgpf_delta_pa}a, the respective posterior distributions are shown.   
Other cases of varying polarization that were reported earlier were GRB 041219A \citep{Gotz_etal_2009}, GRB 100826A \citep{Yonetoku_etal_2011} and GRB 170114A \citep{Zhang_etal_2019} observed by {\it INTEGRAL}, GAP and POLAR respectively.   
%determined at confidence levels of $99.95 \, \%$ ($3.5 \, \sigma$) and $99.83\, \%$ ($3.1 \, \sigma$) respectively.
The low PF, thus, found across the burst can now be understood as an artefact of the temporal change of PA occurring within the burst. The results of the polarization analysis of the three time intervals are listed in Table \ref{t_pol}.

\begin{figure}
\plotone{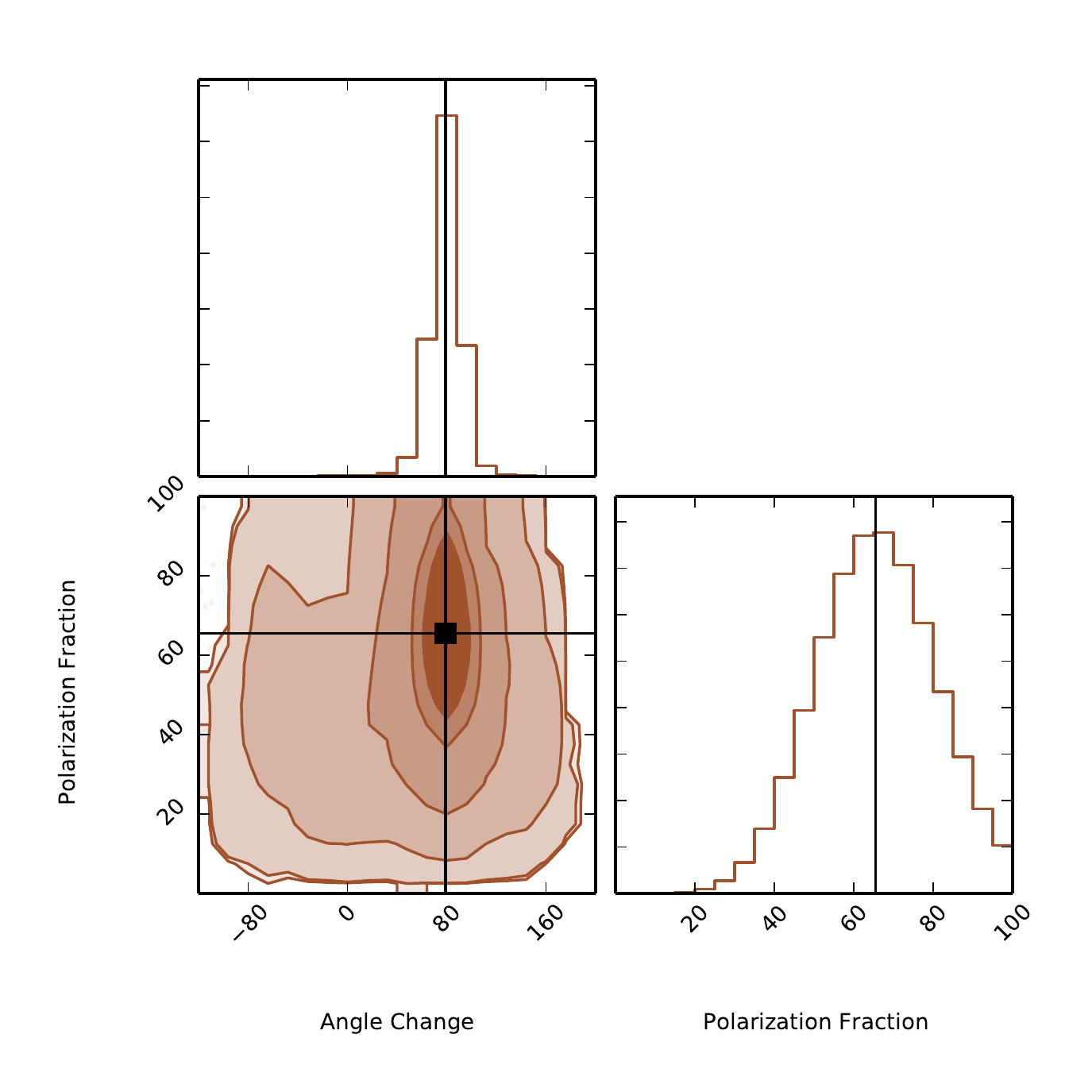}
\caption{Above a 2D histogram plot of the average PF and correspondingly obtained average change in polarization angle, across the burst is shown. Contours corresponding to confidence levels $68\,\%, \, 90\,\%, \, 99.7\,\%, \, 99.99\, \%$, $ 99.99992\, \% \, (5.3\, \sigma)$ and $ 99.99995\, \% \, (5.4\, \sigma)$ are shown. The average values of PF and change in PA are marked on their respective posterior distributions by a black solid line. 
}
\label{avgpf_delta_pa}
\end{figure}

In order to estimate the average PF across the burst:
(a) the first and the third intervals were combined since they had nearly same polarization angles (fourth panel of Figure \ref{lc_whole_pol} c); 
(b) Monte Carlo simulations involving combined fits of this new interval and the second interval with the cosine function were performed. The polarization fraction related parameters A and B of the cosine functions were linked across the intervals, while the polarization angles were kept free. Thus, we found the average polarization fraction and the polarization angles for the new interval and the second interval. In Figure \ref{avgpf_delta_pa}, the posterior distributions of the average PF across the burst and the corresponding change in polarization angle estimated by taking the difference of the PAs of the new interval and interval 2 are shown. The average polarization fraction across the burst is estimated 
%\footnote{The methodology adopted in \citealt{Yonetoku_etal_2011} is used.} 
to be $66^{+26}_{-27} \%$ at $99.99992\% \, (5.3\, \sigma)$ confidence for two parameters of interest as shown in Figure \ref{avgpf_delta_pa}. 
Also, we note that the change in polarization angle estimate ($80^{+17\, \circ}_{-18}$) is consistent
with the average of the change in polarization angles that were found occurring within the burst.
%By taking into account the change in polarization angle, 
Previously, such a high statistically significant polarization was reported for the burst GRB 021206 by \cite{Coburn_boggs_2003}, however, the claim was contested by the analyses done by \cite{Rutledge_Fox_2004} and  \cite{Wigger_etal_2004} subsequently. Recently, POLAR found that the time integrated emission of 5 bright GRBs observed by them, possess the most probable polarization fraction between $4\%$ and $11\%$ at a confidence level of $99.9\, \%$ \citep{Zhang_etal_2019}. 
Till date no other polarization measurement of GRBs reported by BATSE, {\it INTEGRAL}, GAP, {\it AstroSat} and POLAR have obtained a statistical significance greater than $ \sim 99.9\, \%$  \citep{Zhang_etal_2019, Covino_Gotz2016, McConnell2017}.

%In case of GRB160821A, we note that time resolved bins exhibited high polarization. 

\begin{table}
\let\nobreakspace\relax
\begin{small}
\caption{Results of the temporal polarization analysis. \\ $^*$ The average values of $\mu$, PF and PA  from the posterior distribution are reported. 
%The reported errors are $68\%$ confidence intervals for two parameters of interest including the systematic errors.
}
\label{t_pol}
\begin{center}
\renewcommand{\arraystretch}{2.0}
\begin{tabular}{|p{3.0cm}|c|c|c|}
\hline	
Time Intervals (s) & 115-129 & 131-139 & 142-155  \\ 
\hline 
Compton events  & 896 & 1124 & 523 \\
%Compton events (s$^{-1}$)  & 64 & 140.5 & 40.2 \\
Background events (s$^{-1}$) & $20.7 \pm 4.5$ & $20.7\pm 4.5$ & $20.7 \pm 4.5$\\
MDP ($99\%$ confidence) & $40\%$& $32\%$&$57\%$\\
$\mu^{*}$ & 0.289 & 0.250 & 0.243\\
$\mu_{100}$ &0.409 & 0.435 & 0.403  \\
%PF$^*$ &$71^{+26}_{-34}\,\%$ & $58^{+23}_{-24}\%$ &$61^{+33}_{-37}\, \%$ \\
PF$^*$ &$71^{+29}_{-41}\,\%$ & $58^{+29}_{-30}\%$ &$61^{+39}_{-46}\, \%$ \\
PA$^*$ &$110^{+14\, \circ}_{-15}$ &$31^{+12\, \circ}_{-10}$ & $110^{+25\, \circ}_{-26}$\\
Confidence level & $99.8\%$ & $99.97\%$ & $99.1\%$
\\ \hline
\end{tabular}
\end{center}
\end{small}
\end{table}

\section{Spectral analysis}
\begin{figure}
%\plotone{spec_figs.pdf}
\includegraphics[scale=0.70]{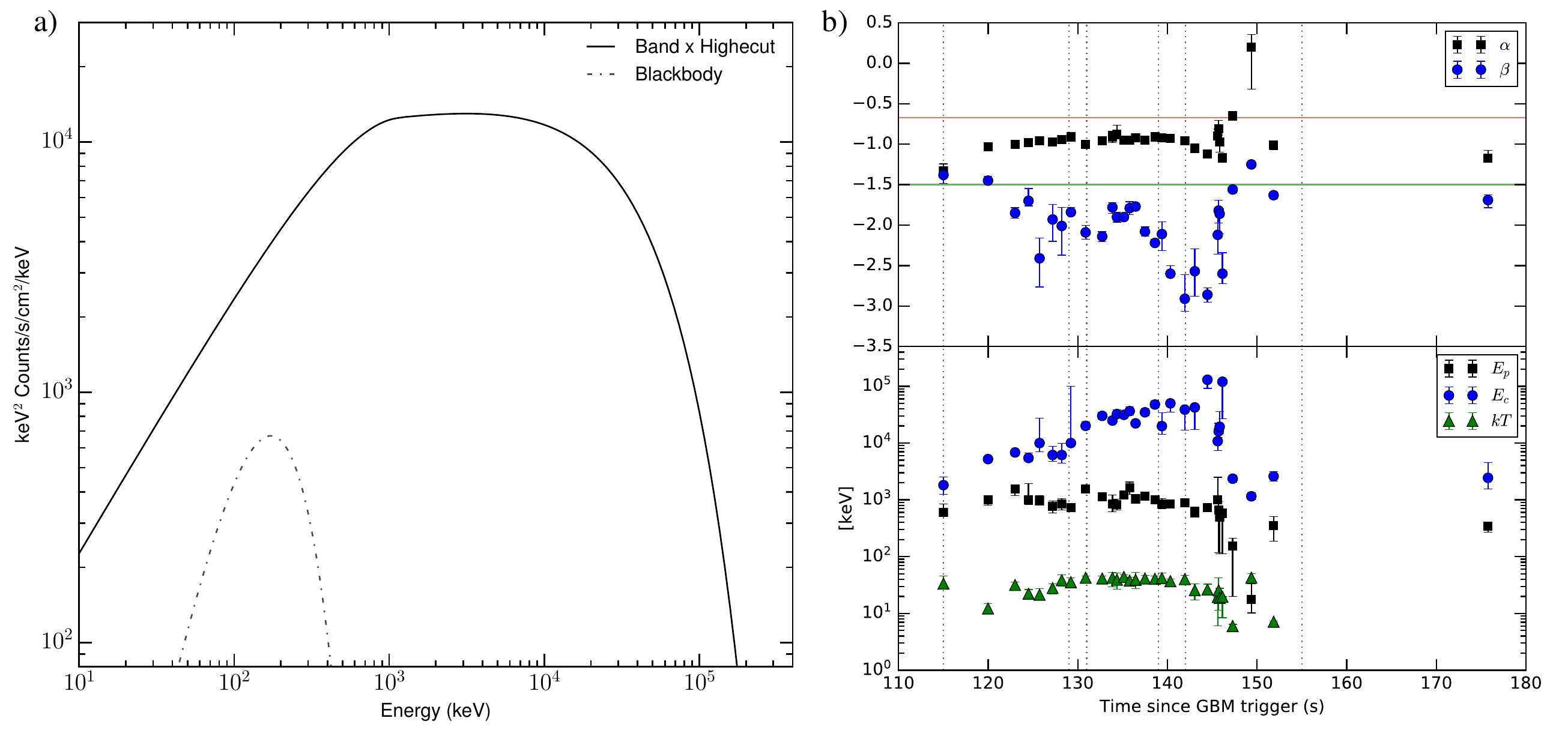}
\caption{a) Figure shows the $\nu F_{\nu}$ plot of the best fit model, 
%Band function only (green solid line), Band function (blue solid line) + blackbody (blue dash dot line) and $f_{BHec}$ 
Band $\times$ Highecut (black solid line) + blackbody (black dash dot line), fitted to the brightest time interval, i.e $134.59 - 135.71$ s.  b) Upper panel shows the time evolution of $\alpha$ (black squares) and $\beta$ (blue circles) of the Band function. The green and red horizontal lines in the upper panel mark the photon index of $\alpha = -1.5$ and $\alpha = -0.67$ corresponding to the fast and slow cooling synchrotron emissions respectively. In the lower panel, the time evolution of  the Band $E_{p}$ (black squares), the high energy cutoff $E_{c}$ (blue circles) and the temperature of the blackbody component k$T$ (green triangles) are shown. The three time intervals of polarization study are shown in black dotted vertical lines across the two panels.}
\label{spec}
\end{figure}

Traditionally, the GRB prompt emission spectrum is modelled using the phenomenological Band function\footnote{Band function is an empirical function consisting of two power laws smoothly joined at a peak.} \citep{Band1993}. 
The time resolved spectral analysis of the main episode of the burst, however, shows significant deviations from the pure Band function in the brightest bins \citep{Preece2014,Vianello_etal_2018}.  The deviation in lower energies is modelled by adding a blackbody (BB) function at $k\rm T \sim 30 \, \rm keV$  and that at higher energies by adding a cutoff at $E_{\rm c} \sim 2 - 50 \, \rm MeV $ (Appendix \ref{spectrum}). 
Thus, the spectrum is best modelled using a blackbody +  Band $\times$ Highecut (Figure \ref{spec}a), where the blackbody can be related to the thermal emission, a relic of the dense fireball formed at the central engine after the explosion, and the rest to the non-thermal emission coming from the optically thin region of the outflow \citep{Guiriec2011,Axelsson2012,Iyyani2013,Burgess2014a,Iyyani_etal_2016,Vianello_etal_2018}. The evolution of the spectral fit parameters are shown in Figure \ref{spec}b.
%and the details of the spectral model are presented in section 4 in supplementary material. 
We note that the low energy part of the spectrum characterized by the low energy power law index, $\alpha$ and the spectral peak, $\rm E_{p}$ are nearly steady throughout at $\sim -0.97$ and $800 \, \rm keV$ respectively. However, the high energy part of the spectrum characterized by the high energy power law index, $\beta$ and cutoff, $\rm E_{c}$ vary significantly such that $\beta$ decreases, whereas $\rm E_{c}$ increases with time and after $T_0 + 140 \, \rm s$, these trends are reversed. This spectral behaviour is consistent with the trend observed in HR reported above. 

\section{Discussion and Summary}
GRBs with long durations, $> 2\, \rm s$, are associated with the death of massive stars. 
A highly collimated outflow (jet) of opening angle $\theta_j$, moving at a relativistic speed (parameterized by Lorentz factor\footnote{$\Gamma = 1/\sqrt{1- (v/c)^2}$ where $v$ is the velocity of the outflow and $c$ is the speed of light.}, $\Gamma$) is produced after the core of the massive star collapses to a black hole (or a magnetar) and begins to accrete the surrounding stellar matter. The radiation emitted from this relativistic outflow is beamed towards the observer within a cone of $1/\Gamma$ which is thus, the visible region around the line of sight. This is referred to as the relativistic aberration/ beaming. 

In a classical fireball model \citep{Meszaros2006,Pe'er2015,Iyyani2018}, where the outflow is non-magnetized, the non-thermal emission is generally expected to be produced in shocks created in the optically thin region above the photosphere from where the thermal emission is expected. Energetic electrons produced in the shocks then cool by processes like synchrotron emission in random magnetic fields generated in the shocks \citep{Ghisellini_Lazzati1999}, or inverse Compton scattering \citep{Lazzati_etal_2004}, both being inherently locally axisymmetric around the outflow's velocity vector. 
For a fixed viewing angle\footnote{$\theta_v$ is the angle between the line of sight of the observer and the jet axis.}, $\theta_v > 0$, in an axially symmetric jet, the polarization vector integrated over the spatially unresolved emitting region, should point either perpendicular to or in the plane containing the axis of the jet and the line of sight of the observer. Thus, the PA is expected to change by $90 ^{\circ}$ when the width of the jet parameterized by $ \Gamma \, \theta_j$ changes. 
In such a case, when the orientation of the polarization vector becomes perpendicular to the observer plane, the polarization fraction is expected to be low, $< 10\%$ \citep{Toma_etal_2009, Granot2003, Ghisellini_Lazzati1999}. However, here we observe that during the  
entire burst, the PF values are $> 15 \, \%$.
%at $68\%$ confidence level of two parameters of interest including systematic error. 
Thus, a change in width of the visible portion of the jet cannot account for the observed change in PA. Inherently axisymmetric emission models referred to above are ruled out.

The total radiative energy\footnote{We find a fluence of $1.3 \pm 0.03 \times 10^{-3}$ erg/cm$^2$ in $10$ keV - $5$ GeV. Fluence is the energy flux integrated over the duration of the burst. We assume a flat Universe ($\Omega_{\lambda} =0.73$ and $H_0 = 71$ km s$^{-1}$ Mpc$^{-1}$). } released in the prompt emission of GRB 160821A is estimated to be $E_{\gamma, \rm iso} = 6.9 \times 10^{53} \, (3.6 \times 10^{55})\, \rm erg$, for a redshift, $z = 0.4 \, (2)$, if the radiation were isotropic. These are among the highest known values for long GRBs \citep{Racusin2011}. Such a high $E_{\gamma, \rm iso}$ suggests that the emission is strongly collimated, with the jet  
pointing towards the observer such that the line of sight lies within the jet cone or just outside the edge of the jet ($\theta_j + 1/\Gamma$). 
In the above scenario, the strong observed polarization can be explained only by synchrotron emission produced in magnetic fields that are highly ordered within the viewing cone of $1/\Gamma$. 

If we assume the observed burst emission is a single emission episode, the observed high polarization along with a change in polarization angle is challenging to be explained in any known physical model. On the other hand, 
it can also be envisaged that the burst emission consists of multiple emission episodes and depending on the dominance of the synchrotron emissions coming from the different regions, a change in polarization angle can happen with time (also see \citealt{Lazzati_Begelman2009}).

Thus, for the first time a conclusive evidence of high and varying linear polarization is detected in a GRB. The observations are extremely constraining and challenging for currently proposed physical models for gamma-ray bursts. This motivates further research into the development of a physical scenario that can explain the observations self consistently.

\newpage 
\bibliographystyle{aasjournal}
\bibliography{ref_160821A}

\section*{Acknowledgments}
We would like to thank Dr Christoffer Lundman, Prof. Pawan Kumar and Dr Santosh Roy for enlightening discussions. This publication uses data from the {\it AstroSat} mission of the Indian Space Research Organisation (ISRO), archived at the Indian Space Science Data Centre (ISSDC). CZT-Imager is built by a consortium of institutes across India, including the Tata Institute of Fundamental Research (TIFR), Mumbai, the Vikram Sarabhai Space Centre, Thiruvananthapuram, ISRO Satellite Centre (ISAC), Bengaluru, Inter University Centre for Astronomy and Astrophysics, Pune, Physical Research Laboratory, Ahmedabad, Space Application Centre, Ahmedabad. 
This research has made use of {\it Fermi} data obtained through High Energy Astrophysics Science Archive Research Center Online Service, provided by the NASA/Goddard Space Flight Center. The Geant4 simulations for this paper were performed using the HPC resources at The Inter-University Centre for Astronomy and Astrophysics (IUCAA)  and Physical Research Laboratory (PRL). \\

\appendix

%\section{Appendix information}
\section{Systematic error estimate in polarization measurement}
\label{sys_err}
Cadmium Zinc Telluride Imager (CZTI) on board AstroSat is an X-ray spectroscopic instrument and is experimentally verified for polarization measurement capability in 100-400 keV energy range for on axis sources \citep{Vadawale_etal_2015}. CZTI functions as a wide field monitor at energies $> 100\, \rm keV$ because the CZTI collimators and other supporting structures become largely transparent above this energy. This enables it to detect GRBs and measure their polarization. 
%The increasing transparency of the CZTI collimators and other supporting structures at energies $> 100$ keV enables it to function as a wide field monitor providing a unique opportunity for GRB detection and polarization measurement.
In order to calculate the modulation factor for off -axis sources, a mass model of AstroSat was developed in Geant4 (version 4.10.02.p02).
There are several possible sources of systematic errors in the measurement of polarization \citep{Chattopadhyay_etal_2017}. Below we list these sources and the error estimates are quoted for the case of GRB 160821A:\\
(i) An uncertainty can be induced in the observed $\mu$ due to different photon propagation paths through the spacecraft structures. This uncertainty is estimated by conducting $\sim 10^4$ Geant4 simulations of this burst with the same spectra and incident direction to produce the same number of observed Compton events. The uncertainty on $\mu$ is thus estimated to be $\sim 8 -16 \% $ according to the different number of Compton events. % for the first and second intervals whereas it is $\sim 17\%$ for the third interval and in case of the whole burst. 
(ii) The selection of background is also expected to induce some systematic error. This was investigated by taking both pre and post GRB background independently as well as in combination, to determine the modulation amplitude. The uncertainty on $\mu$ due to this is found to be $< 1\%$. 
(iii) The effect of localization uncertainty is studied. GRB 160821A is localized at RA = 171.25 and Dec = +42.33 at $+/- 3$ arcmin accuracy \citep{BATrefined_GCN_2016}. The contribution of localization uncertainty on the obtained modulation amplitude and polarization angle is found to $<1 \%$. 
(iv) The uncertainty associated with the spectral model of the GRB is investigated by varying the power law index within its estimated 1 sigma error and we find the associated uncertainty on the observed modulation amplitude to be $< 1\%$. 
(v) Another source of systematic error could be the unequal quantum efficiency of the CZTI pixels. The relative pixel efficiency across the CZTI plane is found to vary within $\leq 5\%$ which produces negligible false modulation amplitude.

In the $\mu_{100}$ estimations, the statistical error is quite small as the simulations are done for a large number ($10^8$) of incoming photons. The systematic errors are those associated with the sources (iii) and (iv) while the uncertainty associated with source (i) is estimated to be $\le 1\%$. The value of $\mu_{100}$ strongly depends on the fitted polarization angle. The uncertainty associated with this in the estimation of PF is taken care of in the Monte Carlo process to obtain the posterior distribution of PF (described in Appendix \ref{posterior}), where for each fitted polarization angle, the corresponding $\mu_{100}$ is used.

%the interpolated value from a table of $\mu_{100}$ values which were estimated using Geant4 simulations of $100 \%$ polarized emission for the same GRB spectra and incoming direction with a discrete grid of polarization angles. 

All these uncertainties are propagated into the reported values of the limits of the $68 \%$ confidence interval of two parameters of interest namely the observed polarization fraction and angle.
%of each time interval.

\section{Monte Carlo Method to obtain posterior distributions of PF and PA}
\label{posterior}
The polarization signature in the GRB is estimated through the non-uniform azimuthal distribution of GRB counts. We calculate the normalized counts for 8 bins whose mid-values correspond to azimuthal angles: 0, 45, 90, 135, 180, 225, 270, 315 degrees. These angles are estimated in anti-clockwise direction when CZTI is viewed from top. %$> 20$. 
%The raw azimuthal distribution obtained after the subtraction of pre and post-backgrounds, is then followed by correction of pixel geometry and off-axis viewing angles. 
The corrected modulation curves are fitted for large number of iterations ($10$ million) using Monte Carlo method for estimating the modulation factor and polarization angle, as follows:\\ 
%We have employed the following randomization method for obtaining the posterior distribution:\\
(a) For each azimuthal angle ($\eta_{\rm i}$), and its corresponding normalized counts ($y_{\rm i}$) and error ($y_{\rm err,i}$), we pick a random value ($y_{\rm ran,i}$), from a normal distribution which is defined by the mean value, $\mu_{\rm mean} = y_{\rm i}$ and the standard deviation, $\sigma_{\rm SD} = y_{\rm err,i}$. A normal distribution is assumed because here, in the case of GRB 160821A, each azimuthal angular bin has over $20$ Compton events \citep{Cash1979}. \\
(b) The A, B and $\phi_0$ parameters of the fitting cosine function given in equation (1), are estimated for each set of randomly drawn $y_{\rm ran,i}$ (where $i$ = 1 to 8) values via least square curve fitting method. For the fitted polarization angle, the corresponding interpolated value of $\mu_{100}$ from a table of $\mu_{100}$ values that were generated for a discrete grid of polarization angles via the Geant4 simulations of $100\%$ polarized emission for the same GRB spectra and incoming direction, is chosen. Thereby, estimating the PF values. \\
(c) The above two steps are repeated for a large number of times ($10^7$), thereby obtaining the various required likelihood distributions of PF and PA. \\
(d) The obtained likelihood is then filtered through the prior condition such as 
%i) The polarization angle can only lie between the curve fit PA\footnote{The PA obtained by fitting the cosine function to the observed azimuthal distribution.} - $90^{\circ}$ and curve fit PA + $90^{\circ}$ and ii) 
the polarization fraction has to lie between 0 and 100 $\%$. The simulation runs which satisfied the prior condition were used to generate the posterior distribution.\\ 
(e) Finally, the respective 2D histogram of posterior distributions of PA and PF are made.

% \section{Estimate of average PF across burst}
% \label{avg_PF}
% In case of GRB160821A, we note that time resolved bins exhibited high polarization as well as a change in polarization angle with time. Thus, in order to estimate the average PF across the burst:
% (a) the first and the third intervals were combined since they had nearly same polarization angles (fourth panel of Figure \ref{lc_whole_pol} c); 
% (b) Monte Carlo simulations involving combined fits of this new interval and the second interval with the cosine function were performed. The polarization fraction related parameters A and B of the cosine functions were linked across the intervals, while the polarization angles were kept free. Thus, we find the average polarization fraction and the
% polarization angles for the new interval and the second interval. In Figure \ref{avgpf_delta_pa}, the posterior distributions of the average PF across the burst and the corresponding change in polarization angle estimated by taking the difference of the PAs of the new interval and interval 2 are shown. We find this change in polarization angle estimate ($80^{+17\, \circ}_{-18}$) to be consistent
% with the average of the change in polarization angles that were found occurring within the burst.

\section{Time resolved spectroscopy}
\label{spectrum}
For analysis of the {\it Fermi} GBM data, three sodium iodide (NaI) detectors with the highest count rates were chosen, namely n6, n7 and n9. 
These detectors observed the GRB at an off-axis angle less than $40^{\circ}$. 
In addition, the Bismuth gallium oxide (BGO) detector BGO 1, which had the strongest detection, was chosen for analysis. LAT data ($>100$ MeV) belonging to the P8$\_$TRANSIENT020E class and its corresponding instrument response were used, whereas, the LAT-LLE spectra (30 - 130 MeV) were obtained using the same method as that for GBM data. 
The spectrum for each {\it Fermi} detector was extracted using the software {\it Fermi} Burst Analysis GUI v 02-01-00p1 (gtburst3 \footnote{https://fermi.gsfc.nasa.gov/ssc/data/analysis/scitools/gtburst.html}). 
The background was obtained by fitting a polynomial function to the data regions before and after the GRB for time intervals, $T_{0}$ - 410 s to $T_{0}$ - 10 s and $T_{0}$ + 210 s to $T_{0}$ + 250 s, respectively.   

A joint spectral analysis of {\it Fermi} GBM along with LAT-LLE and LAT data was carried out in Xspec \citep{Arnaud1996} 12.9.0n software and Pg-stat \citep{Xspec_statistics_2013} statistic was used.  %We obtain the joint \textit{Fermi} detectors spectral fitting results using the Poisson data with Gaussian background statistics, pgstat in Xspec. 
%\cite{Xspec1996} \cite{Xspec_statistics_2013}
For NaI data, energies between $30\, \rm keV - 40\, \rm keV$ corresponding to iodine K-edge and extreme edges such as energies below $10 \, \rm keV$ and above $850 \, \rm keV$ were removed. 
In case of BGO, LAT LLE and LAT, data between energies $300 \, \rm keV -10 \, \rm MeV$, $30 \, \rm MeV - 130 \, \rm MeV$ and $100 \, \rm MeV - 5 \, \rm GeV$ were used respectively. 

For time resolved spectroscopy of the burst, time intervals were defined using Bayesian Blocks (BB) algorithm on the detector (n6) with largest number of counts.  
In the tails of the emission episode, due to low signal to noise ratio, certain blocks were combined to get a larger time interval so that good spectral constraints could be obtained.

For estimating the effective area correction\footnote{An effective area of an instrument translates to its sensitivity, which can be different at different energies and off axis angles from the axis of the instrument. Here in order to do a combined spectral analysis of the data from different instruments like NaI, BGO and LAT, a correction needs to be applied between these instruments keeping one of them as the reference. This is estimated by multiplying a constant, independent of energy, to the normalization of the spectral model chosen for each instrument during the spectral analysis.} for NaI and LAT detectors, the bright time bins were simultaneously fitted with the best fit model i.e blackbody +  Band $\times$ Highecut multiplied with a constant of normalization, while the constant of normalization of the BGO 1 detector was fixed to unity. The relative normalization constants $ 0.97^{+0.01}_{-0.01}$ for n6, $0.92^{+0.01}_{-0.01}$ for n7, $0.94^{+0.01}_{-0.01}$ for n9 and $0.84^{+0.06}_{-0.06}$ for LAT were obtained. Throughout the time resolved spectral analysis, the effective area corrections for the detectors were frozen to these obtained values.

% \subsection{Results}
% In the time resolved spectral analysis {\it (22, 39-41)} of the main episode of the burst, we find significant deviations in the spectrum from the Band function alone fit to the data, characterized by the wavy structures seen in the residuals of the fit, refer Fig. S6 (a, b). 
% The deviation at lower energies is modelled by adding a blackbody function at $k\rm T \sim 30 \, \rm keV$  and that at higher energies is modelled by adding a cutoff at $E_{\rm c} \sim 2 - 50 \, \rm MeV $. 
% The detailed time resolved spectral analysis results for the best fit model, blackbody (BB)  + Band $\times$ Highecut ($f_{BHec}$) {\it (21)}, for the main episode of the burst, is reported in Table S1. %\ref{t3}.  
% %\cite{Abdo2009_080916c, Preece2014, Ackermann2014, Iyyani2018} \cite{Vianello_etal_2018}

\subsection{Statistical Significance test}
\label{sig}

\begin{figure}
\plotone{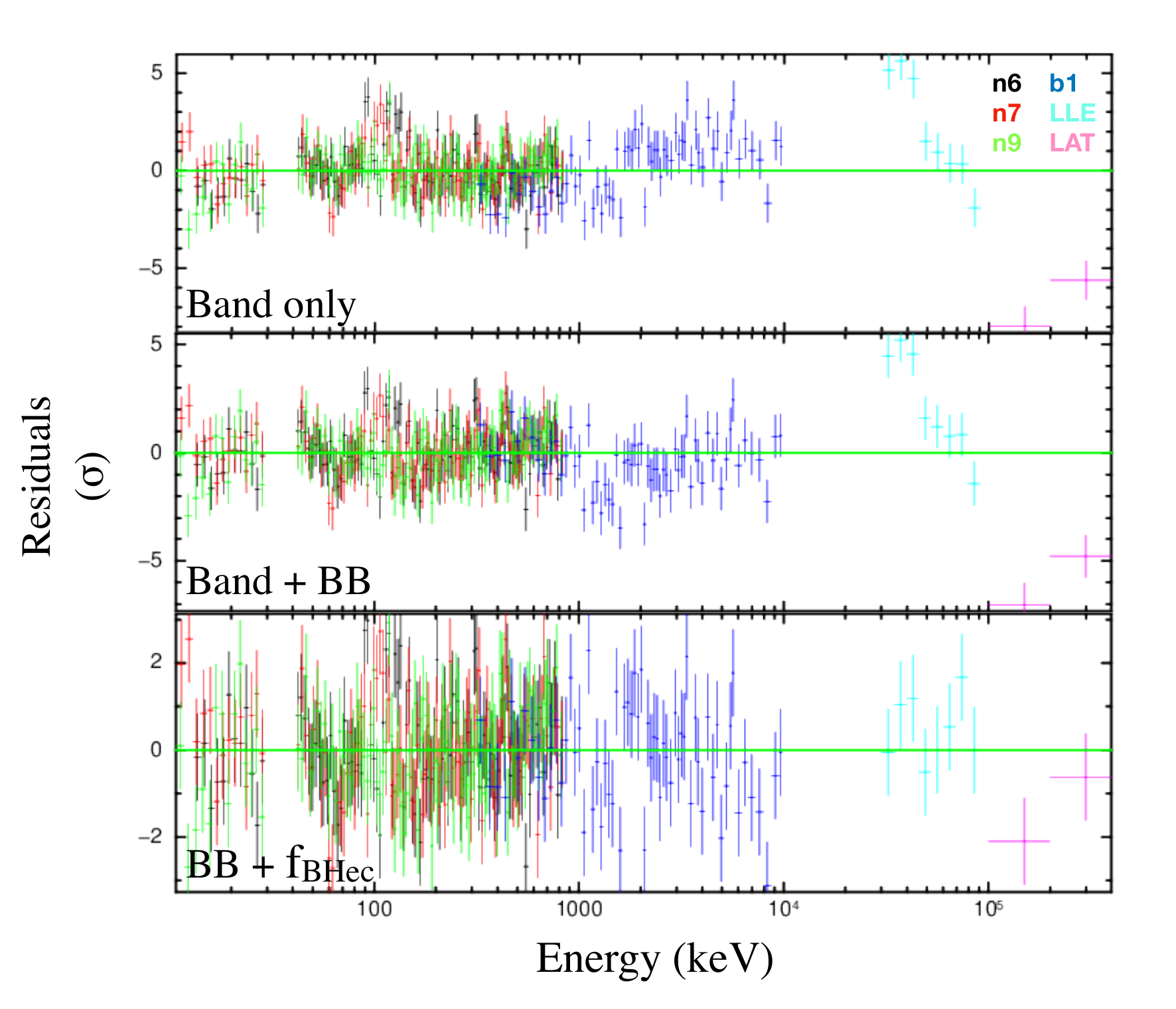}
\caption{The residuals obtained for the different models, Band, Band + BB and BB +$f_{BHec}$,  fits done to the spectral data of the brightest time interval are shown.}
\label{residuals}
\end{figure}

\begin{figure}
\plotone{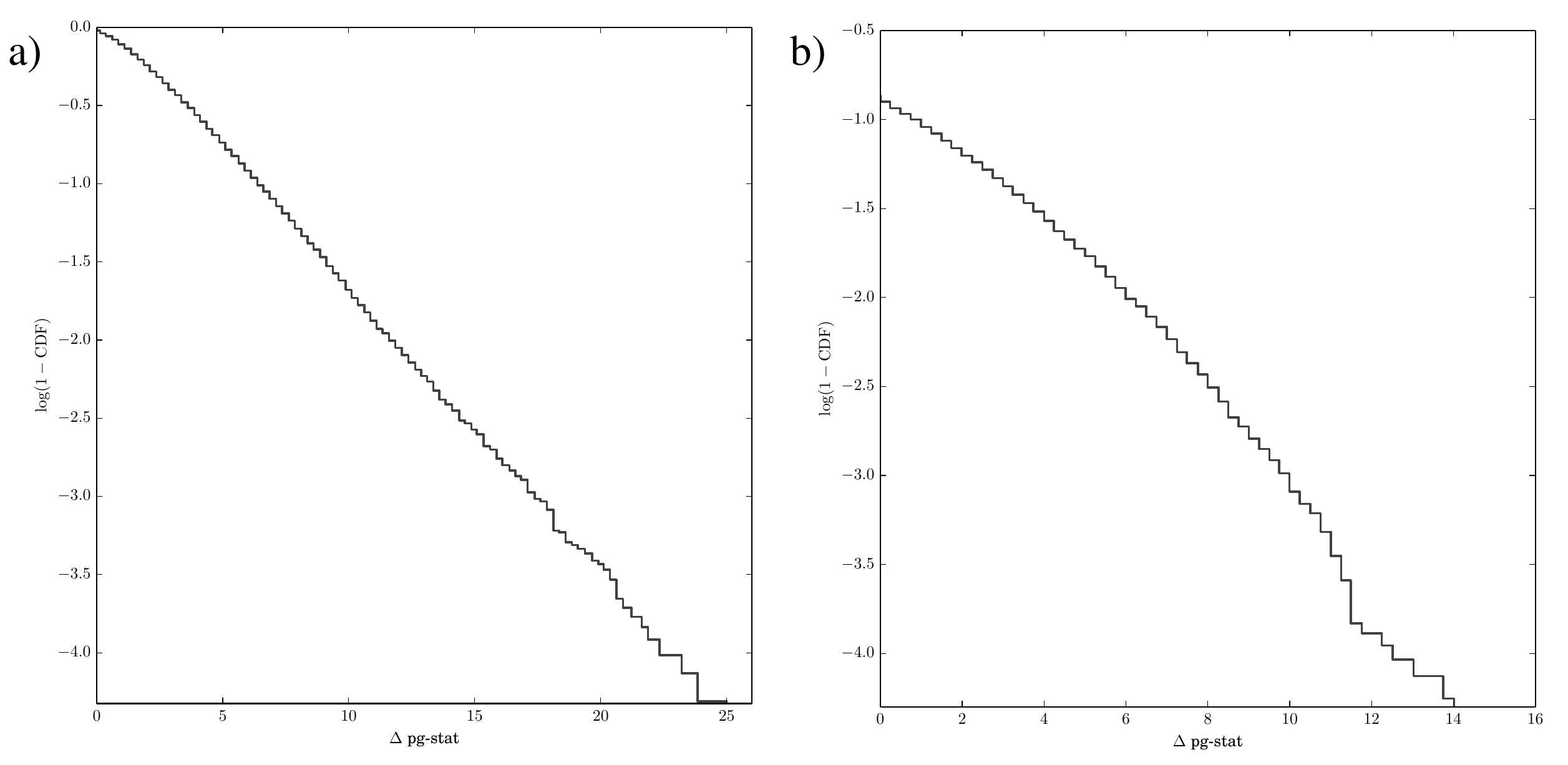}
\caption{The complementary cumulative distribution function (CDF) of the   pg-stat obtained for models a) $f_{BHec}$ function vs blackbody + $f_{BHec}$, and b) blackbody + Band function vs blackbody + $f_{BHec}$ are shown.
}
\label{stats}
\end{figure}

We conduct Monte Carlo simulations in order to ascertain the statistical significance of the deviations in the spectrum from a pure Band function, which have been modelled using a blackbody and a cutoff at lower and higher energies respectively (see the fit residuals in Figure \ref{residuals}). Since this process is computationally intensive here we present the simulation study done for the brightest time bin (134.59 - 135.71 s) only, and adopt the obtained $\Delta$ Pg-stat distribution as a reference to assess the statistical significance of these components in other time bins. \\ 
a)  Blackbody \\
The difference in statistic i.e $\Delta $ Pg-stat, obtained for the model $\rm BB \, +$ Band $\times$ Highecut ($ f_{BHec}$) from $f_{BHec} $ is 24.5. 
We assume the model  $f_{BHec}$ with the best fit model parameters as the null hypothesis (H0) and generate nearly 40,000 realizations and its corresponding background spectra using the {\it fakeit} command in Xspec.  Each of these realizations are then fit with the null hypothesis model $f_{BHec}$ and the alternate hypothesis (H1) model  $\rm BB + f_{BHec}$, and the respective $\Delta $ Pg-stat values are recorded. 
The probability to observe any $\Delta $ Pg-stat of $> 24.5$ is found to be $10^{-4.3}$ (Figure \ref{stats}a) which corresponds to a significance level of   $\sim  4 \,\sigma$.  
\\
b)  Highecut \\
The $\Delta $ Pg-stat, obtained for the model $\rm BB$ $+ f_{BHec}$ from $\rm BB +Band $ is  201. 
In this case, we assume the model  $\rm BB +Band $ with the best fit model parameters as the H0 and generate nearly 54,000 realizations and its corresponding background spectra. The model $\rm BB$ $+ f_{BHec}$ is the H1 and the respective complementary cumulative distribution of $\Delta$ Pg-stat is obtained. 
The probability to obtain $\Delta$ Pg-stat $ = 14 $ is $ 10^{-4.5}$ (Figure \ref{stats}b) which indicates that the probability to obtain any  $\Delta$ Pg-stat $> 201$ is $\ll 10^{-4.5}$   which corresponds to a significance level of $>  4.2 \,\sigma$.  Thus, the addition of highecut to the Band function is highly significant. % $\gg 7 \sigma$.

\end{document}